\documentclass[10pt,aps,prx,twocolumn,notitlepage,showpacs,superscriptaddress]{revtex4-1}
\usepackage{amsthm,amsmath,amsfonts,amssymb,verbatim,color}
\usepackage{graphicx}
\usepackage{subfigure}
\usepackage{bm}
\usepackage{epsfig,slashed}
\usepackage[T1]{fontenc}
\usepackage[colorlinks=true,citecolor=blue,linkcolor=blue,urlcolor=blue]{hyperref}
\usepackage{psfrag}

\begin{document}

\newcommand{\tr}{\mathop{\mathrm{Tr}}}
\newcommand{\bsigma}{\boldsymbol{\sigma}}
\newcommand{\re}{\mathop{\mathrm{Re}}}
\newcommand{\im}{\mathop{\mathrm{Im}}}
\renewcommand{\b}[1]{{\boldsymbol{#1}}}
\newcommand{\diag}{\mathrm{diag}}
\newcommand{\sign}{\mathrm{sign}}
\newcommand{\sgn}{\mathop{\mathrm{sgn}}}
\renewcommand{\c}[1]{\mathcal{#1}}
\renewcommand{\d}{\text{\dj}}

\newcommand{\mb}{\bm}
\newcommand{\ua}{\uparrow}
\newcommand{\da}{\downarrow}
\newcommand{\ra}{\rightarrow}
\newcommand{\la}{\leftarrow}
\newcommand{\mc}{\mathcal}
\newcommand{\bs}{\boldsymbol}
\newcommand{\lra}{\leftrightarrow}
\newcommand{\nn}{\nonumber}
\newcommand{\half}{{\textstyle{\frac{1}{2}}}}
\newcommand{\mf}{\mathfrak}
\newcommand{\MF}{\text{MF}}
\newcommand{\IR}{\text{IR}}
\newcommand{\UV}{\text{UV}}
\newcommand{\be}{\begin{equation}}
\newcommand{\ee}{\end{equation}}

\newcommand{\itt}{\it}
\newcommand{\black}{\textcolor{black}}
\newcommand{\red}{\textcolor{black}}
\newcommand{\ph}{\phantom}
\newcommand{\redd}{\textcolor{black}}
\newcommand{\reddd}{\textcolor{black}}

\DeclareGraphicsExtensions{.png}

\title{Enhancement of Superconducting $T_c$ due to the Spin-orbit Interaction}

\author{Joel Hutchinson}
\affiliation{Department of Physics, University of Alberta, Edmonton, Alberta T6G 2E1, Canada}
\author{J. E.  Hirsch}
\affiliation{Department of Physics, University of California, San Diego, CA, USA, 92093-0319}
\author{Frank Marsiglio}
\affiliation{Department of Physics, University of Alberta, Edmonton, Alberta T6G 2E1, Canada}

\date\today

\begin{abstract}
We calculate the superconducting $T_c$ for a system which experiences Rashba spin-orbit interactions. Contrary to the usual case where the electron-electron interaction is assumed to be  wave vector-independent, where superconductivity is suppressed by the spin-orbit interaction (except for a small region at low electron or hole densities), we find an enhancement of the superconducting transition temperature when we include a correlated hopping interaction between electrons. This interaction   originates in the expansion of atomic orbitals due to electron-electron
repulsion and gives rise to superconductivity only at high electron (low hole) densities. When superconductivity results from
this interaction it is  enhanced by spin-orbit coupling, in spite of a suppression of the density of states. The degree of electron-hole asymmetry about the Fermi surface is also enhanced.
\end{abstract}

\pacs{
}

\maketitle

\section{Introduction}

\red{Spin-orbit coupling is prevalent in condensed matter systems, and can have a profound impact
on the properties of metals and insulators~\cite{winkler, bauer12}, not just at surfaces, but in the bulk.
For superconductivity, the spin-orbit interaction was invoked immediately following BCS~\cite{bardeen57},
mainly to address discrepancies in the Knight shift measurements~\cite{reif57,androes59} 
and the predictions from BCS theory~\cite{ferrell59,martin59,schrieffer59,anderson59,abrikosov62}. 
}

\redd{More recently, %(2001),
as more superconductors with a crystal structure that lacked a centre of inversion symmetry were
becoming common, %Gor'kov and Rashba \cite{gorkov01} revived 
this discussion was revived \cite{edel1989, gorkov01}, utilizing the Rashba model of 
spin-orbit coupling \cite{rashba1959, rashba1960}, and once again focussing attention on the non-zero Knight shift at low temperature. These papers also explicitly identified the novel feature in these
superconductors: a mixed singlet-triplet state, which was implicit all along since spin had been identified in the early work
as {\it not} a good quantum number.} The impact on thermodynamic properties (including the superconducting critical temperature,
$T_c$) was not really considered. Indeed, in Anderson's initial treatment of this problem \cite{anderson59}, he essentially
repeated the arguments made in his more famous "Dirty Superconductors" paper \cite{anderson59_dirty}, but now for
spin-orbit coupling, with the implication that for weak spin-orbit coupling $T_c$ would be unaffected.

\red{A few years later it was pointed out that in principle a large enhancement in $T_c$ could occur, because of an enhancement
in the electronic density of states in the low density region, due to an effective "dimensionality reduction" \cite{cappelluti07}.
However, as we further demonstrate below, this enhancement is confined to a rather narrow electron density window, and the
overall scale of $T_c$ is low for a weakly coupled system.
}

\red{An interesting general question remains, which is the impact of the Rashba spin-orbit interaction on superconducting $T_c$ in the presence of different types of pairing interactions. 
\reddd{Some calculations have been recently performed in Ref. [\onlinecite{ptok2018}] for the extended Hubbard model. }The generic short-range attractive interaction 
(e.g. the attractive Hubbard model) already results in a mixed singlet-triplet state due to the spin-orbit interaction. 
\reddd{However, as we will show (and also found in Ref. [\onlinecite{ptok2018}]), in that case the spin-orbit interaction suppresses superconductivity.} In this paper we
include a specific off-diagonal term in the interaction, previously considered by two of us \cite{hirsch1989}, in the context of cuprate
superconductivity. This interaction is noteworthy in that it has the form of an off-diagonal matrix element of the Coulomb
interaction between electrons in Wannier orbitals, rather than a diagonal matrix element representing   a density-density repulsion or attraction.  
As explained at length previously \cite{hirsch2001,hirsch2000,hirsch2002_qmc}, the so-called ``correlated hopping'' interaction
arises inevitably because the many-body electron wave functions make significant adjustments to minimize the energy associated with
Coulombic repulsions.
}

\red{In the following section we will introduce the model, and briefly discuss some important one-electron properties. These
determine the appropriate basis with which we consider the pairing interaction, the so-called Rashba basis. We follow the usual
BCS description for the pairing state; this leads to a simple parameterization of the wave vector dependence of the order parameter, 
in the presence of spin-orbit coupling. We then present results for $T_c$ as a function of the various interaction strengths and as
a function of the electron density. In general, with the correlated hopping interaction present, spin-orbit coupling leads to a significant
enhancement of superconducting $T_c$. We then end with a summary.
}

\section{Tight-binding Hamiltonian\protect\red{, including correlated hopping and spin-orbit coupling}}

\red{As described in earlier work \cite{hirsch1989,micnas1990}, a tight-binding model that includes both the on-site Hubbard ``$U$''
interaction and the correlated hopping term, ``$\Delta t$,'' is}
\begin{eqnarray}\label{eq:H1}
H_{\rm Mod} &=&-t\sum_{\langle ij \rangle \atop \sigma} (c_{i\sigma}^{\dagger
}c_{j\sigma}+c_{j\sigma}^{\dagger }c_{i\sigma})  + U\sum_i n_{i\uparrow} n_{i\downarrow} \\
&& + \Delta t \sum_{\langle  ij \rangle \atop \sigma} \bigl( c_{i\sigma}^{\dagger
}c_{j\sigma}+c_{j\sigma}^{\dagger }c_{i\sigma} \bigr) \bigl( n_{i -\sigma} + n_{j -\sigma} \bigr).
\end{eqnarray}
Here, $c^\dagger_{i\sigma}$ creates an electron on site $i$ with spin $\sigma=\uparrow$,$\downarrow$, and $\langle ij\rangle$ means that we only consider hopping between nearest neighbour sites $i,j$. In what follows we will assume a square lattice.
The tight-binding parameters are the hopping integral $t$, the on-site repulsion $U$ and the correlated hopping parameter $\Delta t$, described in detail in \cite{hirsch1989}. \red{Briefly, this term represents the fact that electrons will hop with an altered hopping parameter
when other electrons are nearby. It was considered by Hubbard in his original publication on the Hubbard model \cite{hubbard1963a},
and then dropped as he focussed on the on-site interaction alone.}

%{\color{red} The following should maybe be moved to introduction and/or appendix} 

We add to this Hamiltonian a Rashba spin-orbit coupling term~\cite{rashba1959}. 
\redd{Such a term is generic for systems that either
lack inversion symmetry~\cite{bychkov1984} or experience some Fermi surface instability~\cite{wu2007}}, while maintaining time-reversal symmetry as well as a uniaxial symmetry. For a square lattice, the most generic spin-dependent quadratic hopping term, restricted to nearest neighbours, is
\be\label{eq:rashba1}
H_{\rm SO}=\sum_{i\alpha\beta}(c^\dagger_{i\alpha}\vec{a}\cdot\vec{\sigma}^{\alpha\beta}c_{i+\hat{x},\beta}+c^\dagger_{i\alpha}\vec{b}\cdot\vec{\sigma}^{\alpha\beta}c_{i+\hat{y},\beta}+h.c.).
\ee
The uniaxial symmetry to be enforced is a rotation by $\pi/2$ about the $\hat{z}$ axis through each site $i$. Applying such a rotation to this term using $R_{\phi}c_{j\alpha}R^{-1}_{\phi}=e^{-i\alpha\phi}c_{R^{-1}_{\phi}j,\alpha}$ gives
\begin{eqnarray}
H_{\rm SO}&=&\sum_{i\alpha\beta}e^{i(\alpha-\beta)\pi/4}(c^\dagger_{i+\hat{y},\alpha}\vec{a}\cdot\vec{\sigma}^{\alpha\beta}c_{i,\beta}+c^\dagger_{i\alpha}\vec{b}\cdot\vec{\sigma}^{\alpha\beta}c_{i+\hat{x},\beta})\nonumber\\
&&+h.c..
\end{eqnarray}
Matching this to \eqref{eq:rashba1} restricts the values of $\vec{a}$ and $\vec{b}$ to be
\redd{\begin{eqnarray}
a_x&=&0=b_y\\
a_y&\equiv&-iV_{\rm SO}=-b_x.
\end{eqnarray}}
Thus the Rashba hopping term on the direct lattice is
\be\label{eq:rashbahop}
H_{\rm SO}=V_{\rm SO}\sum_{i\alpha\beta}\bigg(ic^\dagger_{i\alpha}\sigma_x^{\alpha\beta}c_{i+\hat{y},\beta}-ic^\dagger_{i\alpha}\sigma^{\alpha\beta}_yc_{i+\hat{x},\beta}\bigg)+h.c.,
\ee
where $V_{\rm SO}$ parameterizes the Rashba spin-orbit coupling. In general, this parameter depends on the atomic spin-orbit coupling and \red{on} the details of the band structure, and should be determined from experiment or ab initio studies~\cite{winkler, petersen2000}. The largest values of $V_{\rm SO}$ typically occur at surfaces or interfaces (e.g. BiTeI has $V_{\rm SO}/t\sim0.8$~\cite{fu2013}). It is important to recognize that in the tight-binding picture, a Rashba term should be present whenever there is inversion asymmetry in the site point group (with some preserved uniaxial symmetry)~\cite{Zhang2014}. This means that even quasi-two-dimensional materials whose crystal structure is centrosymmetric can have bulk Rashba spin-splitting if there is polarity in any given plane. This is true, for example in YBCO, where the Yttrium and Barium ions on opposite sides of the copper oxide planes produce a local electric dipole moment. For the cuprates, the Rashba parameter has been estimated to be $V_{\rm SO}/t\sim0.008$~\cite{Edelstein2004}. Larger spin splittings ($V_{\rm SO}/t\sim0.04$) can be found in the LaAlO$_3$/SrTiO$_3$ interface, which supports a superconducting 2D electron gas, though the magnitude of this splitting is still under debate~\cite{shalom2010, zhong2013}. 

In a single-band model where the correlated hopping interaction $\Delta t$ arises simply from an off-diagonal matrix element
of the Coulomb interaction between neighboring Wannier orbitals, as discussed by Hubbard \cite{hubbard1963a} and 
others \cite{kivelson1987,micnas1990}, 
the spin-orbit interaction would not be expected to modify the interaction terms in the Hamiltonian.
Instead, within the `dynamic Hubbard model'  \cite{hirsch2001} the interaction $\Delta t$ arises from
the modification of the on-site electron wavefunction when another electron occupies the site, due to Coulomb repulsion.
\redd{This effect is modeled by the site Hamiltonian \cite{hirsch2001}
\be
H_i=\omega a_i^\dagger a_i+[U+g\omega (a_i^\dagger+a_i)]n_{i\uparrow}n_{i\downarrow}
\ee
where the boson creation and annihilation operators $a_i^\dagger, a_i$ describe the electronic excitations of an
electron when a second electron is added to the orbital. A generalized Lang-Firsov transformation
\cite{lang1963, hirsch2001}
\be
c_{i\sigma}=e^{g(a_i^\dagger-a_i)\tilde{n}_{i,-\sigma}}\tilde{c}_{i\sigma}\equiv X_{i\sigma}\tilde{c}_{i\sigma}
\ee
relates the original fermion operators $c_{i\sigma}$ to new fermion quasiparticle operators $\tilde{c}_{i\sigma}$ that both destroy the
electron at the site and change the state of the boson so that the boson field follows the fermion motion.
Since $X_{i\sigma}^\dagger=X_{i\sigma}^{-1}$, the transformation preserves fermion anticommutation relations.
To obtain a low-energy effective Hamiltonian we consider only ground-state to ground-state transitions of the boson field,
and in this approximation the relation Eq. (9) becomes \cite{hirsch2001}}
\be
{c}_{i\sigma}^\dagger=[1-(1-S)\tilde{n}_{i,-\sigma}]\tilde{c}_{i\sigma}^\dagger
\ee
\redd{where $S=e^{-g^2/2}$. The on-site repulsion $U$ is lowered to $U_{eff}=U-\omega g^2$, and bilinear terms in
fermion operators at different sites transform as follows:
\begin{eqnarray}
c_{i\sigma}^\dagger c_{j\sigma '}&=&
\tilde{c}_{i\sigma}^\dagger \tilde{c}_{j\sigma '}
[1-(1-S)(\tilde{n}_{i,-\sigma}+\tilde{n}_{j,-\sigma '})\nonumber\\
&&+(1-S)^2 \tilde{n}_{i,-\sigma} \tilde{n}_{j,-\sigma'}]  .
\end{eqnarray}
We will be interested in the regime where the band is close to full. The coefficient in the parenthesis of Eq. (11)
when the occupations are such that $\tilde{n}_{i,-\sigma}+\tilde{n}_{j,-\sigma '}=1$ is $S$, and when
$\tilde{n}_{i,-\sigma}+\tilde{n}_{j,-\sigma '}=2$ is $S^2$. Their difference is
\be
S-S^2=S(1-S)\equiv \frac{\Delta t}{t}
\ee
which defines the correlated hopping $\Delta t$ in this model. The term involving
$\Delta t$ in Eq. (2) then results from replacing the bare operators $c_{i\sigma}$ by the 
quasiparticle operators $\tilde{c}_{i\sigma}$ in the hopping term
(and renaming the quasiparticle operators 
$\tilde{c}_{i\sigma}\rightarrow c_{i\sigma}$), and discarding terms involving six fermion operators that will
be unimportant for low hole concentration \cite{hirsch2000}.} Similarly the spin-orbit interaction term
  \eqref{eq:rashbahop} is modified to
\begin{eqnarray}
H_{\rm SO}=iV_{\rm SO}\sum_{i, \alpha \beta}\bigg(c_{i,\alpha }^{\dagger }\sigma _{x}^{\alpha \beta }c_{i+
\hat{y},\beta }[1-\frac{\Delta t}{t}(n_{i,\beta}+n_{i+\hat{y},\alpha})]\nonumber\\
-c_{i,\alpha }^{\dagger }\sigma _{y}^{\alpha \beta }c_{i+\hat{
x},\beta }[1-\frac{\Delta t}{t}(n_{i,\beta}+n_{i+\hat{x},\alpha})]\bigg)+h.c.,\nonumber\\
\label{eq:H_SOC}
\end{eqnarray}
so that the full Hamiltonian of our model is $H_{\rm Mod}+H_{\rm SO}-\mu\sum_{i,\sigma}n_{i\sigma}$, where $\mu$ is the chemical
potential. 

In the usual BCS fashion, we Fourier transform this Hamiltonian and eliminate interactions between pairs 
\red{with} finite momentum to obtain a reduced Hamiltonian
\begin{eqnarray}\label{eq:Hred}
H&=&\sum_{\b{k},\sigma}(\epsilon_\b{k}-\mu)c^\dagger_{\b{k}\sigma}c_{\b{k}\sigma}-2V_{\rm SO}\sum_{\b{k}}\bigg(\sin k_y(c^\dagger_{\b{k}\uparrow}c_{\b{k}\downarrow}+c^\dagger_{\b{k}\downarrow}c_{\b{k}\uparrow})\nonumber\\
&&+i\sin k_x(c^\dagger_{\b{k}\uparrow}c_{\b{k}\downarrow}-c^\dagger_{\b{k}\downarrow}c_{\b{k}\uparrow})\bigg)\nonumber\\
&&+\frac{1}{N}\sum_{\b{k}}V^0(\b{k},\b{k}')c^\dagger_{\b{k}\uparrow}c^\dagger_{-\b{k}\downarrow}c_{-\b{k}'\downarrow}c_{\b{k}'\uparrow}\nonumber\\
&&+\frac{1}{N}\sum_{\mathbf{kk}'}\sum_{\alpha\beta}\bigg(V^{\rm R}_{\alpha\beta}(\b{k}')c^\dagger_{\mathbf{k}\alpha}c^\dagger_{-\mathbf{k}\beta}c_{\mathbf{k}'\alpha}c_{-\mathbf{k}'\alpha}\nonumber\\
&&+V^{\rm R}_{\alpha\beta}(\b{k})c^\dagger_{\mathbf{k}\beta}c^\dagger_{-\mathbf{k}\beta}c_{\mathbf{k}'\beta}c_{-\mathbf{k}'\alpha}\bigg),
\end{eqnarray}
where $\epsilon_\mathbf{k}\equiv-2t(\cos k_x +\cos k_y)$ and $V^0(\b{k},\b{k}')\equiv U-2\frac{\Delta t}{t}(\epsilon_\b{k}+\epsilon_{\b{k}'})$ are the dispersion and interaction in the absence of spin-orbit coupling. The correlated hopping and Rashba terms are coupled via the interaction \redd{$V^{\rm R}_{\alpha\beta}(\b{k})\equiv 2V_{\rm SO}\frac{\Delta t}{t}(\sin k_x\sigma^y_{\alpha\beta}+\sin k_y\sigma^x_{\alpha\beta}).$} Throughout this paper, we work in units where the lattice parameter is unity.

\begin{figure}[t]
	\centering
	\includegraphics[width=0.8\columnwidth]{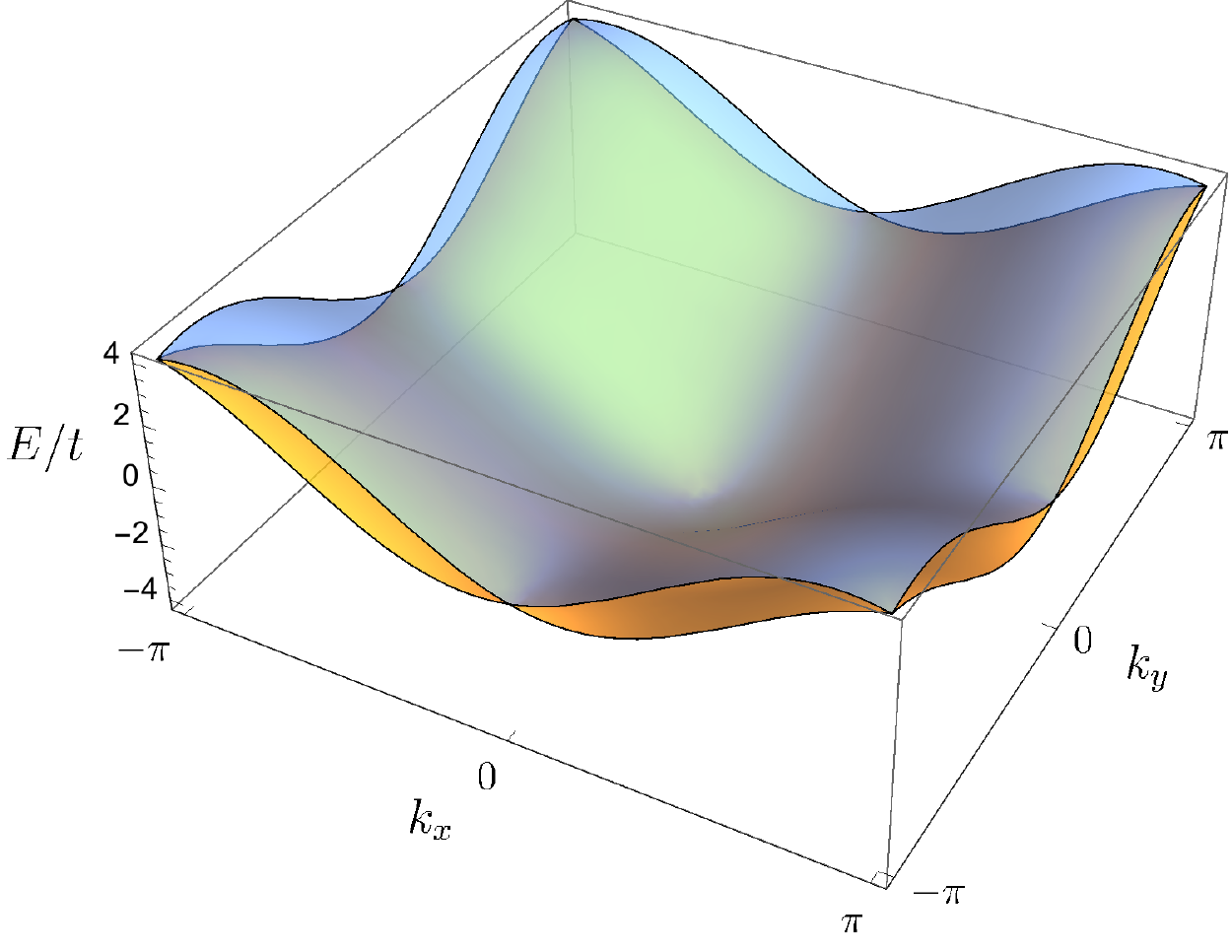}
	\includegraphics[width=0.8\columnwidth]{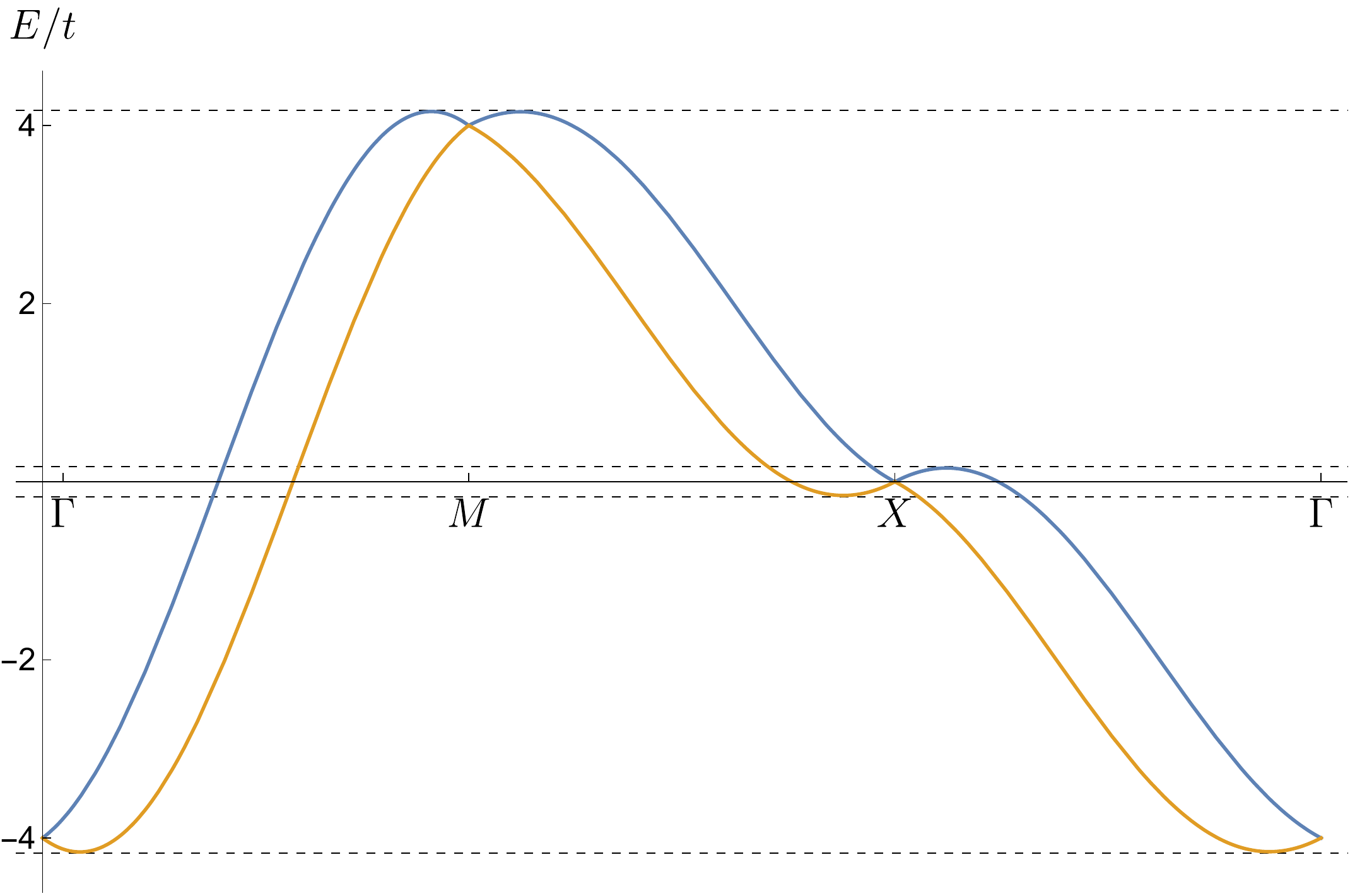}
\caption{Free particle Rashba spectrum on a square lattice with $V_{\rm SO}=0.4t$. The blue and orange bands represent the $s=-1$ and $s=+1$ helicity bands respectively. The dashed lines in the bottom figure show the locations of Van Hove singularities.}\label{fig:spec}
\end{figure}

\begin{figure}[t]
	\centering
	\includegraphics[width=0.8\columnwidth]{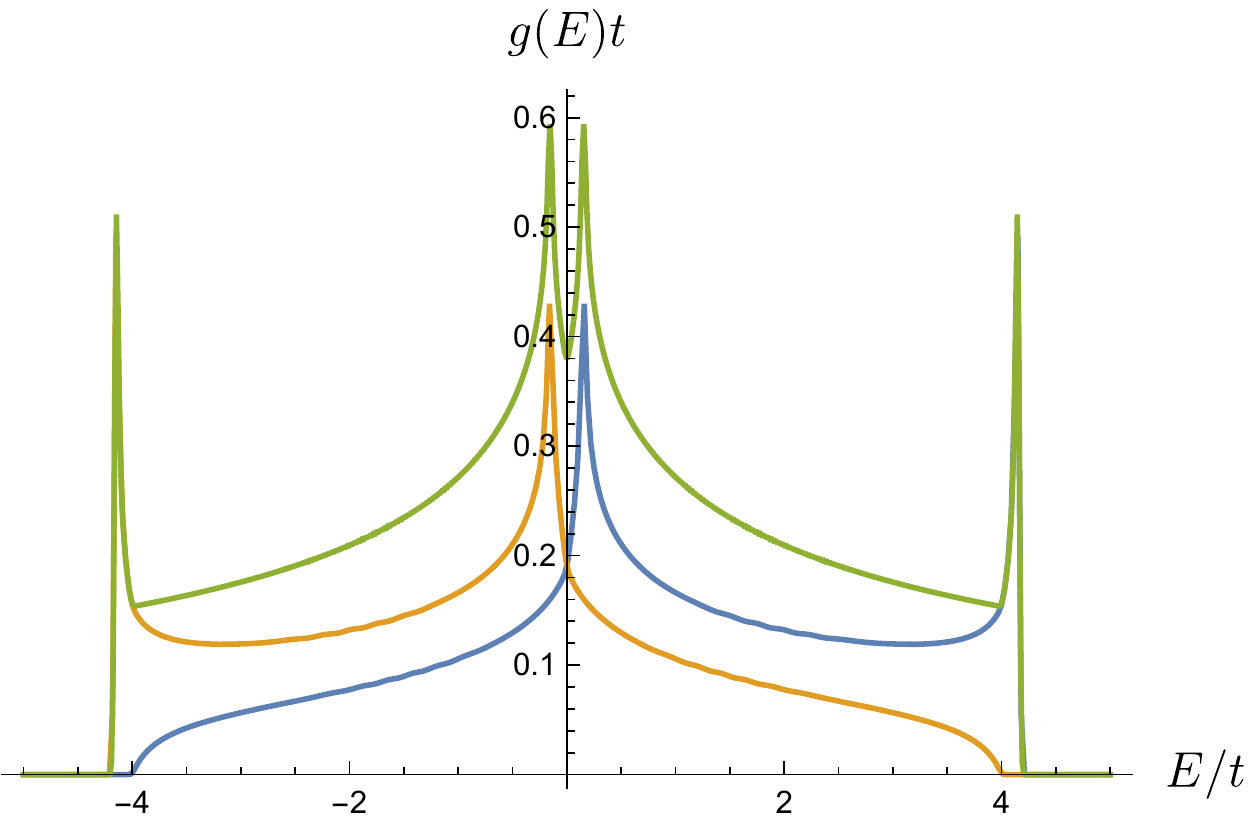}
\caption{Non-interacting single particle Rashba density of states on a square lattice for the lower helicity band (orange), upper helicity band (blue), and the two combined (green). Here we have set $V_{\rm SO}=0.4t$.}
\label{fig:dos}
\end{figure}

The first two lines above constitute the non-interacting Hamiltonian which is diagonalized in the Rashba basis to produce a spectrum
\be
\epsilon_{\b{k}s}=\epsilon_\b{k}-2sV_{\rm SO}\sqrt{\sin^2k_x+\sin^2k_y},
\ee
where $s=\pm1$ represent two helicity branches, with corresponding eigenvectors 
\be\label{eq:helicity}
c^\dagger_{\mathbf{k}s}=\frac{1}{\sqrt{2}}(c^\dagger_{\mathbf{k}\uparrow}+se^{i\theta(\mathbf{k})}c^\dagger_{\mathbf{k}\downarrow}).
\ee
Here we have defined the phase factor
\be
e^{i\theta(\mathbf{k})}\equiv\frac{\sin k_y-i\sin k_x}{\sqrt{\sin^2 k_x +\sin^2 k_y}},
\ee
which governs the mixing of spin-up and spin-down components for eigenstates of the non-interacting Hamiltonian. This mixing ensures that pairs are always formed in a mixed singlet-triplet state. An example of the non-interacting spectrum as well as the density of states is shown in Figures \ref{fig:spec} and \ref{fig:dos} respectively. The density of states is determined by numerically integrating 
\begin{eqnarray}
g(E)&=&4\sum_s\int_0^\pi\frac{dk_x}{2\pi}\int_0^\pi\frac{dk_y}{2\pi}\delta(E-\epsilon_{\b{k}s})\nonumber\\
&=&\lim_{\sigma\rightarrow 0}\frac{1}{\pi^{3/2}\sigma t}\sum_s\int_0^\pi dk_x\int_0^\pi dk_ye^{-(E/t-\epsilon_{\b{k}s}/t)^2/\sigma^2}.\nonumber\\
\end{eqnarray}
\red{Potentially important details concerning van Hove singularities, etc. are carefully derived in Ref.~[\onlinecite{li2011,li2012}].
In particular, while a one-dimensional-like square-root singularity arises at the bottom of the band for parabolic 
dispersion \cite{cappelluti07}, in a tight-binding model the density of states is a constant at the bottom of the band and has a singularity
very close to the bottom of the band where there is a saddle point. Ref.~[\onlinecite{li2011}] makes it clear that for
weak values of $V_{\rm SO}$ the saddle point energy, $E_{\rm sad} = -2t \left[ 1 + \sqrt{1 + (V_{\rm SO}/t)^2} \right]$, is very close to the
minimum energy, $E_{\rm min} = -4t \left[ 1 + \sqrt{V^2_{\rm SO}/(2t^2)} \right]$. Hence, this small separation is not even visible
in Fig.~\ref{fig:dos}.}

%fig. 1

%fig. 2

In the conventional BCS programme, the next step would be to restrict the Hamiltonian \eqref{eq:Hred} to interactions between singlet pairs. In view of the Rashba spin-mixing, however, it is clear that this would not capture the right pairing physics and that it is natural to consider pairs within the same helicity band at zero total momentum. \redd{Allowing for interband pairing in the absence of a magnetic field would be akin to considering FFLO states where pairs have finite total momentum. At zero field, the pairing is expected to be intraband~\cite{loder2013, weng2016}.} Indeed if one follows the prescription of time-reversed pairing due to Anderson~\cite{anderson59}, then $c_{\mathbf{k}s}$ should be matched with its time-reversed partner $-se^{i\theta(\b{k})}c_{\mathbf{-k}s}$. Upon transforming \eqref{eq:Hred} to the helicity basis and retaining only interaction terms involving intra-band pairs, we are left with the effective Hamiltonian
\begin{eqnarray}
H&=&\sum_{\mathbf{k}s}(\epsilon_{\mathbf{k}s}-\mu)c^\dagger_{\mathbf{k}s}c_{\mathbf{k}s}\nonumber\\
&&+\frac{1}{4N}\sum_{\mathbf{kk'}}\sum_{ss'}V_{ss'}(\mathbf{k,k'})c^\dagger_{\mathbf{k}s}c^\dagger_{-\mathbf{k}s}c_{-\mathbf{k'}s'}c_{\mathbf{k'}s'},
\end{eqnarray}
where
\begin{eqnarray}
V_{ss'}(\mathbf{k,k'})&=&s'e^{i\theta(\b{k}')}se^{-i\theta(\b{k})}\bigg(U+8\Delta t(s_\b{k}+s_{\b{k}'})\nonumber\\
&&+4V_{\rm SO}\frac{\Delta t}{t}[s'\sqrt{\sin^2k_x'+\sin^2k_y'}\nonumber\\
&&+s\sqrt{\sin^2k_x+\sin^2k_y}]\bigg).\label{eq:int}
\end{eqnarray}
Here we have defined $s_\b{k}\equiv\frac{1}{2}(\cos k_x+\cos k_y)$.%$\cos\theta_\b{k}\equiv{\rm Re}(e^{i\theta(\b{k})})$, $\sin\theta_\b{k}\equiv{\rm Im}(e^{i\theta(\b{k})})$, and 
%\begin{itemize}
%\item Introduce Rashba SOC and the $V_{SO}\Delta t$ interaction.
%\item Fourier transform and go to the Rashba basis.
%\item Figure: Illustrate non-interacting spectrum, indicating van Hove singularities in the DOS. %(Need to think about what range of $V_{\rm SO}$ we are interested in. For cuprates, this is very small, for LaO/STO, Ce$_3$Pt$_3$Si or UIr, this is much higher~\cite{Edelstein2004, loder2013, sergienko2004}).
%\item Discussion of the reduced Hamiltonian with interactions between intraband pairs. Write out full interaction (without nearest-neighbour repulsion $V$). 

%Point out that this is the appropriate way to follow Anderson's time-reversed pairing prescription. In particular, the time-reversed pair of $c_{-\mathbf{k}s}$ is $se^{i\theta_k}c_{\mathbf{k}s}$. The pairing mean field will then be $b_{\mathbf{k}s}=se^{i\theta_k}\langle c_{-\mathbf{k}s}c_{\mathbf{k}s}\rangle$. This phase factor is important, it means that the gap function will take the form $\Delta_{\mathbf{k}s}=se^{-i\theta_k}\bar{\Delta}_{\mathbf{k}s}$, where $\bar{\Delta}_{\mathbf{k}s}$ is the piece that transforms under the irreducible representations of the lattice. This is nicely explained in reference~\cite{sergienko2004}.
%\end{itemize}

%%%%%%%%%%%%%%%%%%%%%%%%%%%%%%%%%%%
\section{Mean Field Theory}
We now study the effective Hamiltonian within mean field theory. We choose a pairing mean field of time-reversed electron pairs. As discussed above, this is represented in the helicity basis as $b_{\b{k}s}\equiv se^{i\theta(\b{k})}\langle c_{-\b{k}s}c_{\b{k}s}\rangle$. Writing $c_{-\b{k}s}c_{\b{k}s}=se^{-i\theta(\b{k})}b_{\b{k}s}+\delta c_{\b{k}s}$ and neglecting terms of order $(\delta c_{\b{k}s})^2$, we get the mean field Hamiltonian
\begin{eqnarray}
H_{\rm MF}&=&\sum_{\mathbf{k}s}(\epsilon_{\mathbf{k}s}-\mu)c^\dagger_{\mathbf{k}s}c_{\mathbf{k}s}-\frac{1}{2}\sum_{\mathbf{k}s}\Delta_{\mathbf{k}s}^*c_{-\mathbf{k}s}c_{\mathbf{k}s}\nonumber\\
&&-\frac{1}{2}\sum_{\mathbf{k}s}\Delta_{\mathbf{k}s}c_{\mathbf{k}s}^\dagger c_{\mathbf{-k}s}^\dagger
+\frac{1}{2}\sum_{\mathbf{k}s}\Delta_{\mathbf{k}s}se^{i\theta(\b{k})}b^*_{\mathbf{k}s},\nonumber\\
\end{eqnarray}
where we have defined the gap parameter as 
\be
\Delta_{\mathbf{k}s}\equiv -\frac{1}{2N}\sum_{\mathbf{k'}s'}V_{ss'}(\mathbf{k,k'})s'e^{-i\theta(\b{k'})}b_{\mathbf{k'}s'}.\label{eq:DeltaDef}
\ee
Note that this gap function may be written as $\Delta_{\mathbf{k}s}=se^{-i\theta_k}\bar{\Delta}_{\mathbf{k}s}$, where $\bar{\Delta}_{\mathbf{k}s}$ transforms under an irreducible representation of the lattice point group (in this model the trivial representation of the dihedral group $D_8$). The unusual phase factor $se^{-i\theta_k}$ that is local in k-space is a feature of spin-orbit coupling and is discussed in Ref.~\cite{sergienko2004}.

The mean field Hamiltonian may be diagonalized by means of the Bogoliubov transformation
\be
c_{\b{k}s}=u^*_{\b{k}s}\hat\alpha_{\b{k}s}-se^{-i\theta(\b{k})}v_{\b{k}s}\hat{\alpha}^\dagger_{-\b{k}s},
\ee
where the coefficients $u_{\b{k}s}$, $v_{\b{k}s}$ are chosen to satisfy $u_{\b{k}s}=u_{-\b{k}s}$, $v_{\b{k}s}=v_{-\b{k}s}$, and $|u_{\bs{k}s}|^2+|v_{\b{k}s}|^2=1$. It is readily found that the values of these parameters that diagonalize the Hamiltonian are given by the equations
\begin{eqnarray}
|v_{\b{k}s}|^2&=&\frac{1}{2}(1-(\epsilon_{\b{k}s}-\mu)/E_{\b{k}s})\\
|u_{\b{k}s}|^2&=&\frac{1}{2}(1+(\epsilon_{\b{k}s}-\mu)/E_{\b{k}s})\\
u_{\b{k}s}v_{\b{k}s}^*&=&-\frac{\bar{\Delta}^*_{\b{k}s}}{2E_{\b{k}s}},
\end{eqnarray}
%\redd{Joel: the minus sign in the last equation is unconventional --- can it be removed through a simple change in some definitions???}\par
where $E_{\b{k}s}\equiv\sqrt{(\epsilon_{\b{k}s}-\mu)^2+|\Delta_{\b{k}s}|^2}$. The final mean field Hamiltonian then reads
\be
H_{\rm{MF}}=\sum_{\b{k}s}E_{\b{k}s}\hat{\alpha}_{\b{k}s}^\dagger\hat{\alpha}_{\b{k}s}+E_g,
\ee
where the ground state energy is given by
%\be
%E_g=\frac{1}{2}\sum_{\b{k}s}((\epsilon_{\b{k}s}-\mu)-E_{\b{k}s}+\Delta_{\b{k}s}se^{i\theta(\b{k})}b_{\b{k}s}^*).
%\ee
\be
E_g=\frac{1}{2}\sum_{\b{k}s}\left[(\epsilon_{\b{k}s}-\mu)-E_{\b{k}s}+\red{\bar{\Delta}_{\b{k}s} b_{\b{k}s}^*}\right]. %\Delta_{\b{k}s}se^{i\theta(\b{k})}b_{\b{k}s}^*)
\ee

In terms of the new fermionic quasiparticle operators, we have
\be
b_{\b{k}s}=u_{\b{k}s}^*v_{\b{k}s}(2\langle\hat{\alpha}_{\b{k}s}^\dagger\hat{\alpha}_{\b{k}s}\rangle-1),
\ee
which means the gap function must satisfy the finite temperature self-consistency condition
\be
\Delta_{\b{k}s}=-\frac{1}{2N}\sum_{\b{k}'s'}V_{ss'}(\b{k},\b{k}')\frac{\Delta_{\b{k}'s'}}{2E_{\b{k}'s'}}(1-2f(E_{\b{k}'s'})),
\ee
where $f(E)$ is the Fermi function.
%\begin{itemize}
%\item Definition of gap parameter as above.
%\item Mean field Hamiltonian
%\item Self-consistent gap equation.
%\end{itemize}

%%%%%%%%%%%%%%%%%%%%%%%%%%%%%%%%%%%%%%%%%%%%%%%%%%%%%
\section{Gap Equations}
\begin{figure}[t]
	\centering
	\includegraphics[width=0.99\columnwidth]{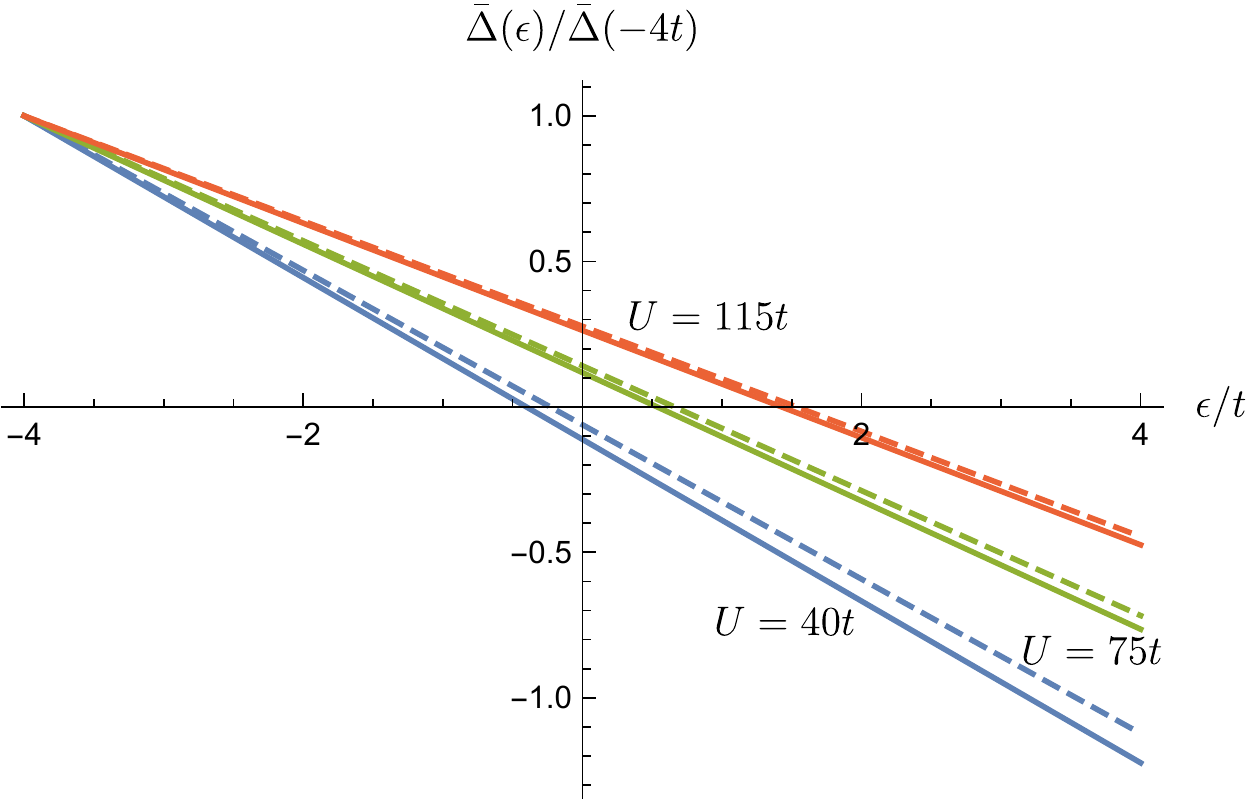}
\caption{Low temperature gap dependence on kinetic energy for various values of $U$. Here we have set $\Delta t=4.5t$, $n=1.875$, $k_BT=0.01t$. The solid lines correspond to $V_{\rm SO}=0.5$. The dashed lines \red{shows the result for} $V_{\rm SO}=0$. A slightly larger value for the absolute value of the slope indicates that $V_{\rm SO}$ increases the asymmetry around the Fermi level.}\label{fig:DelvsEps}
\end{figure}

The self-consistency condition determines the ansatz for $\bar{\Delta}_{\b{k}s}$:
\be
\bar{\Delta}_{\b{k}s}=\Delta^0+\Delta^ss_{\b{k}}+\Delta^{x-y}s\sqrt{\sin^2k_x+\sin^2k_y}.
\ee
Note that while the gap parameter definitively has the (extended) s-wave symmetry of the lattice, it will always be mixed singlet-triplet, unlike in conventional BCS theory. With this ansatz, the self-consistency condition yields three coupled equations
\begin{eqnarray}
\Delta^0&=&-\frac{1}{2N}\sum_{\b{k'}s'}\bigg(U+8\Delta ts_{\b{k'}}\nonumber\\
&&+4\frac{\Delta t}{t}V_{\rm SO}s'\sqrt{\sin^2k_x'+\sin^2k_y'}\bigg)\nonumber\\
&&\times g(E_{\b{k'}s'})\bar{\Delta}_{\b{k}'s'}\label{eq:del0}\\%[(\Delta^0+\Delta^ss_{\b{k'}})+\Delta^{x-y}s'\sqrt{\sin^2k_x'+\sin^2k_y'}]\nonumber\\
\Delta^s&=&-\frac{1}{2N}\sum_{\b{k'}s'}8\Delta tg(E_{\b{k'}s'})\bar{\Delta}_{\b{k}'s'}\label{eq:dels}\\%[(\Delta^0+\Delta^ss_{\b{k'}})\nonumber\\
%&&+\Delta^{x-y}s'\sqrt{\sin^2k_x'+\sin^2k_y'}]\nonumber\\
\Delta^{x-y}&=&-\frac{1}{2N}\sum_{\b{k'}s'}4\frac{\Delta t}{t}V_{\rm SO}g(E_{\b{k'}s'})\bar{\Delta}_{\b{k}'s'}\label{eq:delxmy}%[s'(\Delta^0+\Delta^ss_{\b{k'}})\nonumber\\
%&&+\Delta^{x-y}\sqrt{\sin^2k_x'+\sin^2k_y'}],\nonumber\\
\end{eqnarray}
where we have defined $g(E)\equiv\frac{1}{2E}(1-2f(E))$.
Equations \eqref{eq:dels} and \eqref{eq:delxmy} reveal that there are in fact only two independent parameters since
\be
\Delta^{x-y}=\frac{V_{\rm SO}}{2t}\Delta^s.
\ee
This also means that the gap function is a linear function of the kinetic energy, since
\begin{eqnarray}
\bar{\Delta}_{\b{k}s}&=&\Delta^0+\Delta^s\bigg(s_\b{k}+\frac{sV_{\rm SO}}{2t}\sqrt{\sin^2k_x+\sin^2k_y}\bigg)\\
&=&\Delta^0-\Delta^s\epsilon_{\b{k}s}/\red{(4t)}.
\end{eqnarray}
This energy dependence is in stark contrast with that of the constant gap used in conventional BCS theory. In the context of electron tunnelling, it will cause an energy dependence in the conductance, independent of the density of states. This is seen as an asymmetry in the tunnelling current for bias voltages of different sign. This asymmetry is slightly enhanced by spin-orbit coupling, though the enhancement diminishes with increasing $U$, as seen in Fig. \ref{fig:DelvsEps}.

%fig. 3

The linearized version of the self-consistency equations is a $3\times3$ determinant equation that determines the critical temperature. Note that in the limit of no spin-orbit coupling, $\Delta^{x-y}$ vanishes, and this reduces to two coupled equations which have been solved in Ref.~\cite{hirsch1989}. Due to the presence of the chemical potential in the Fermi function we must simultaneously solve the number equation
\be
n=1-\frac{1}{2N}\sum_{\b{k}'s'}\frac{(\epsilon_{\b{k}'s'}-\mu)}{E_{\b{k'}s'}}\tanh(\beta E_{\b{k}'s'}/2).
\ee
The determinant and number equations are solved together iteratively. \red{That is, we iterate over temperatures until the determinant equation is satisfied, and for each temperature, the chemical potential is found from the number equation. The same strategy is used to solve for the gap function below $T_c$.}  %using Ridder's method~\cite{ridders1979}.
%\be
%\begin{pmatrix}
%1+UI_4+8\Delta tI_5+4\frac{\Delta t}{t}V_{\rm SO}I_6\; & UI_5+8\Delta tI_7+4\frac{\Delta t}{t}V_{\rm SO}I_8\; & UI_6+8\Delta tI_8+4\frac{\Delta t}{t}V_{\rm SO}I_9\\
%8\Delta t I_4 & 1+8\Delta t I_5 & 8\Delta t I_6\\
%4\frac{\Delta t}{t}V_{\rm SO}I_4 & 4\frac{\Delta t}{t}V_{\rm SO}I_5 & 1+4\frac{\Delta t}{t}V_{\rm SO}I_6  
%\end{pmatrix}
%=0.
%\ee
%\begin{itemize}
%\item Ansatz for $\bar{\Delta}_{\mathbf{k}s}$, which includes $\Delta^0$, $\Delta^s$, and $\Delta^{x-y}$. Explain that all of these terms have s-wave symmetry, but the full order parameter will always be mixed singlet-triplet by the nature of the spin-orbit coupling.
%\item Derive the $3\times3$ determinant equation for $T_c$.
%\item Note that in the limit of no SOC, this decouples into $2\times2$ and $1\times1$ equation.
%\item The gap and number equations are solved simultaneously, iteratively, using Ridder's method.
%end{itemize}

%%%%%%%%%%%%%%%%%%%%%%%%%%%%%%%%%%%%%%%%%%%%%%%%%%%%%
\section{Results}
\begin{figure}[t]
	\centering
	\includegraphics[width=0.99\columnwidth]{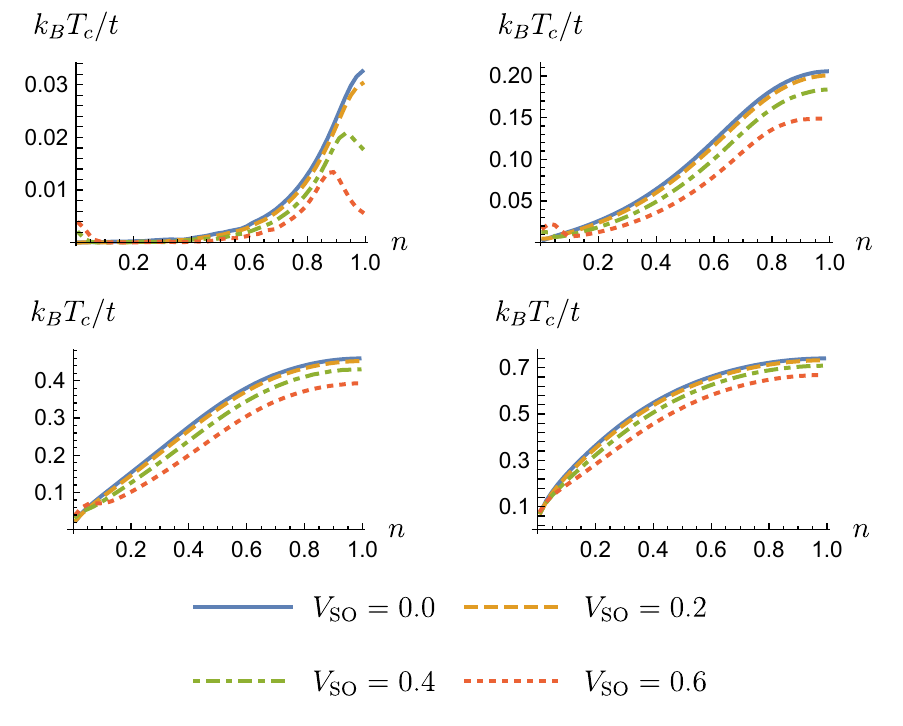}
\caption{Critical temperature as a function of electron density for various values of the spin-orbit coupling with $\Delta t=0$. $U=-t$ (top left), $U=-2t$ (top right), $U=-3t$ (bottom left), and $U=-4t$ (bottom right). By particle-hole symmetry, the plot above half-filling is a reflection of this plot\red{, i.e. $T_c(2-n) = T_c(n)$ for $0 < n < 1$}.}
\label{fig:Tcattract}
\end{figure}

Figure \ref{fig:Tcattract} shows the critical temperature $T_c$ as a function of electron density for the attractive Hubbard model with correlated hopping turned off ($U<0$, $\Delta t=0$). We see that except at very low \red{($n \rightarrow 0$)} and high 
\red{($n \rightarrow 2$)} densities, increasing the spin-orbit coupling has the effect of decreasing the critical temperature. This is understood by noting that the available phase space for intra-band pairs is reduced by the presence of spin-orbit coupling except at the bottom and top of the band where all states are of the same helicity and the density of states becomes singular. Recall that the density of states is lower for the $s=+$ ($s=-$) band in the electon (hole) doped part of the band. Indeed, half-filling, which would have the highest $T_c$ in the absence of spin-orbit coupling shows a dip due to the minimum \red{(see Fig.~\ref{fig:dos})}  in the density of states.

%fig. 4

%fig. 5
\begin{figure}[t]
	\centering
	\includegraphics[width=0.99\columnwidth]{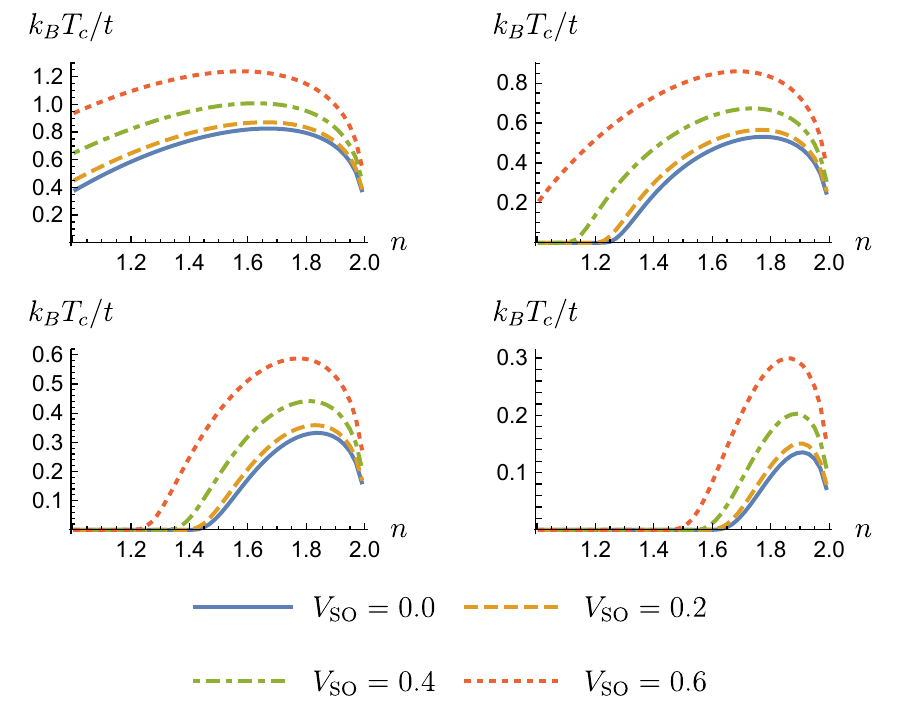}
\caption{Critical temperature as a function of electron density for various values of the spin-orbit coupling with $\Delta t=4.5t$. \redd{ $U=90t$ (left), $U=115t$ (right).}}\label{fig:Tcmod}
\end{figure}
%fig. 6
\begin{figure}[t]
	\centering
	\includegraphics[width=0.99\columnwidth]{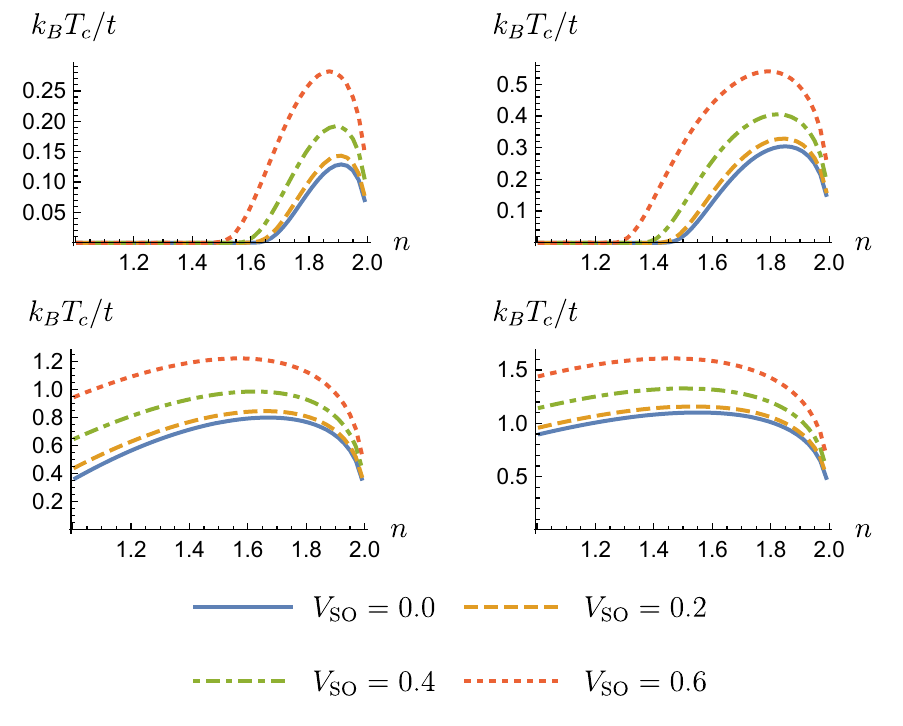}
\caption{Critical temperature as a function of electron density for various values of the spin-orbit coupling with $U=75t$. \redd{$\Delta t=3.5t$ (left), $\Delta t=4t$ (right).}}\label{fig:Tcmod2}
\end{figure}

\begin{figure}[t]
	\centering
	\includegraphics[width=0.99\columnwidth]{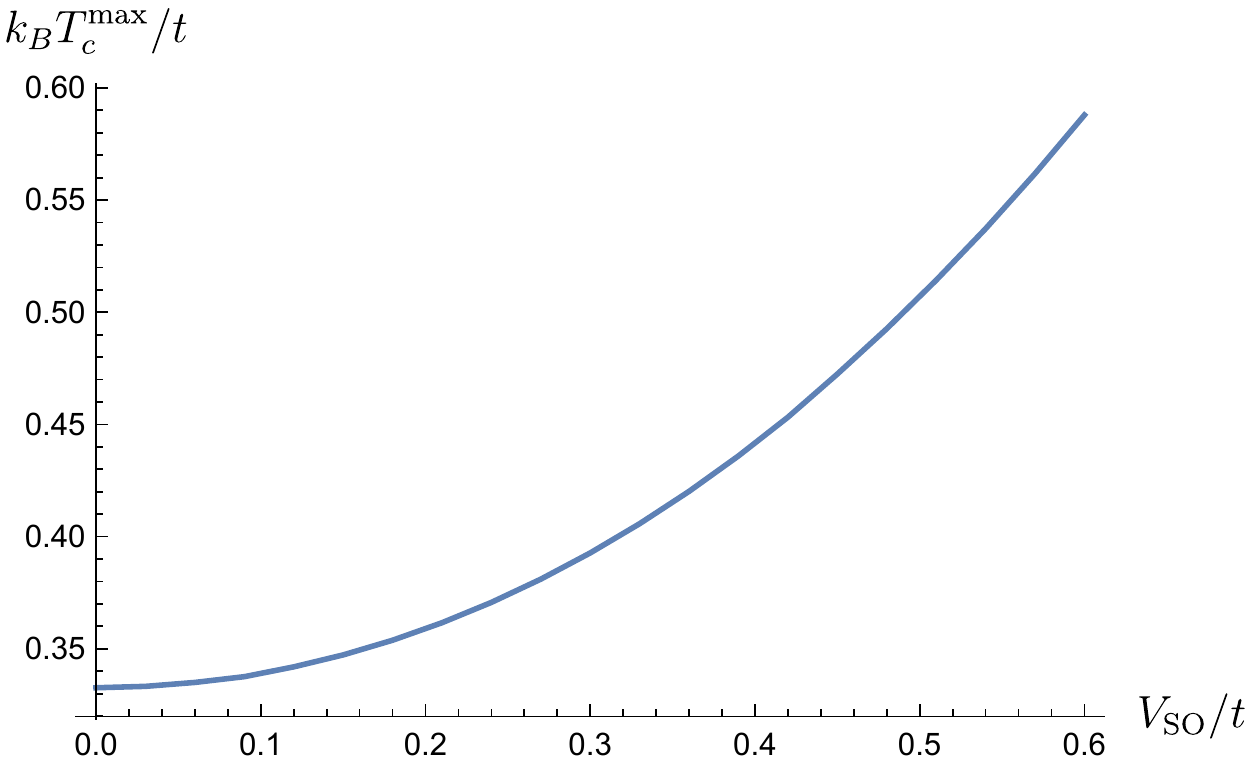}
\caption{Maximum critical temperature as a function of spin-orbit coupling with $U=90t$, $\Delta t=4.5t$.}
\label{fig:Tcmax}
\end{figure}

Figures \ref{fig:Tcmod} and \ref{fig:Tcmod2} show a very different effect. Here correlated hopping has been turned on ($U>0$, $\Delta t\neq0$). This breaks particle-hole symmetry, and we see that the Rashba spin-orbit coupling and correlated hopping cooperate to enhance the critical temperature in the high electron \red{(low hole)} density regime. This too follows from the single-particle density of states. The spin-orbit coupling and correlated hopping couple to produce an effective interaction whose sign matches the sign of the helicity band [see the last two lines of Eq.~\eqref{eq:int}]. At high electron densities, the $s=+$ density of states is suppressed, and the $s=-$ density of states increases towards the singularity at the top of the band. Thus, the pair interaction becomes dominantly attractive and its magnitude increases with $V_{\rm SO}$ and $\Delta t$. In fact the maximum value of the critical temperature shows a quadratic dependence on the spin-orbit coupling as seen in figure \ref{fig:Tcmax}.
%fig. 7

\begin{figure}[t]
	\centering
	\includegraphics[width=0.99\columnwidth]{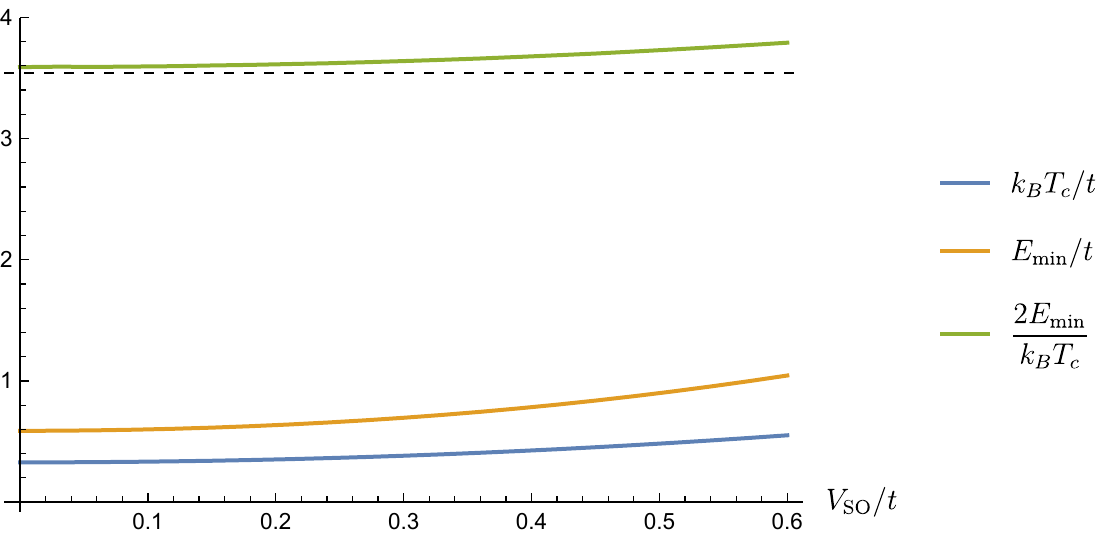}
\caption{Critical temperature, $E_{\rm min}$, and the corresponding gap ratio at low temperature as a function of spin-orbit coupling. Here we have set $U=90t$, $\Delta t=4.5t$, $n=1.875$, $k_BT=0.01t$. The dashed line shows the conventional BCS value.}\label{fig:gapratio}
\end{figure}

The gap and number equations are solved at finite temperature as well. We can check the gap ratio as well, but due to the energy dependence of the gap, it is more appropriate to use the minimum value of the excitation energy. This occurs when
%\be
%\epsilon_{\b{k}s}=\frac{\Delta^0\Delta^s/4+\mu}{1+(\Delta^s/4)^2},
%\ee 
\red{
\be
\epsilon_{\b{k}s}=\frac{\mu + \Delta^0{ \Delta^s \over 4t}}{1+({\Delta^s \over 4t})^2},
\ee
}
at which point the excitation energy is
%\be
%E_{\rm min}=\frac{|\Delta^0-\mu\Delta^s/4|}{\sqrt{1+(\Delta^s/4)^2}}.
%\ee
\red{
\be
E_{\rm min}=\frac{|\Delta^0-\mu {\Delta^s \over 4t}|}{\sqrt{1+({\Delta^s \over 4t})^2}}.
\ee
}
This value is plotted in figure \ref{fig:gapratio} along with the gap ratio. We see that \red{for these parameter values} 
the spin-orbit coupling introduces very little deviation from the BCS value.
%The dependence of $\Delta^0$ on the spin-orbit coupling is shown in figure \ref{fig:del0vsVso}.
%fig. 8

%\begin{itemize}
%\item Figure: $T_c$ vs $n$ for the attractive Hubbard model ($U<0$, $\Delta t=0$) for a variety of $V_{\rm SO}$ (and a variety of $U$?)
%\item Point out that at ultra-low densities (of electrons and holes), there is enhanced $T_c$ due to the singular DOS.
%\item Explain that at intermediate densities $T_c$ is suppressed due to the smaller phase space for upper helicity pairs.
%\item Maybe we should have a figure with $T_c$ vs $V_{\rm SO}$ at fixed density?
%\item Figure: $T_c$ vs $n$ for the correlated hopping model ($U>0$, $\Delta t\neq0$) for a variety of $V_{\rm SO}$ (and a variety of $U$, $\Delta t$?) 
%\item Note the generic rise in $T_c$ at low hole densities due to the attractive interaction $8V_{\rm SO}\frac{\Delta t}{t}s'\sqrt{\sin^2k_x+\sin^2k_y}+...$ where $s'=-1$ for the upper helicity band.
%\item Maybe another figure with $T_c$ vs $V_{\rm SO}$ at fixed density?
%\end{itemize}

%%%%%%%%%%%%%%%%%%%%%%%%%%%%%%%%%%%%%%%%%%%%%%%%%%%%%
%\section{Finite temperature?}
%Do we want to include anything below $T_c$? The discussion would be straight forward, since there are no competing symmetry phases.

%%%%%%%%%%%%%%%%%%%%%%%%%%%%%%%%%%%%%%%%%%%%%%%%%%%%%
\phantom{aaa}

\section{Conclusion}
We have shown that within a 2D tight-binding model on a square lattice, correlated hopping and Rashba spin-orbit coupling work together to enhance the critical temperature of superconductivity, even with significant repulsive on-site interactions. This is in contrast to the Rashba model with attractive on-site interactions and no correlated hopping, where the spin-orbit coupling inhibits superconductivity. The analysis was done within a mean field treatment of the model assuming Cooper pairs to form within the same helicity band of the non-interacting Rashba spectrum. 

The enhancement is strongest in the high electron-density regime. This is relevant for the cuprates at low hole doping, where the oxygen p-band is nearly full. Rashba spin-splitting is expected to be present in many cuprates, though its magnitude is likely much smaller than the values considered in this paper. 

The superconducting gap for this model is thermodynamically similar to the gap in conventional BCS theory in many regards, except that the broken particle-hole symmetry of our model will produce a tunnelling asymmetry in a metal-superconductor junction. 

It is interesting to look at the symmetry of the gap function as well. The gap in this model has an extended s-wave symmetry, but we should note that if nearest neighbour repulsion were added to our model, the gap symmetries will be enriched by the presence of two additional d-wave phases. It should be noted that these are not the symmetries of the full gap function, but rather the part that transforms under irreducible representations of the lattice point group. In particular, the gap carries an additional complex phase due to the spin-orbit coupling. It would be interesting to see if this phase is observable.
\newline
%\begin{itemize}
%\item Mention that the inclusion of nearest neighbour repulsion will give rise to additional p and d-wave phases. 
%\end{itemize}

%   \begin{figure}[t]
%	\centering
%	\includegraphics[width=\columnwidth]{kspace.pdf} %=9cm,angle=0
%	\includegraphics[width=\columnwidth]{spectrum.pdf} %=9cm,angle=0
%\caption{(a) k-space contours and (b) low-energy spectrum for a single Rashba electron. The shaded region shows the allowed virtual transitions with $|k-k_0|<\tilde\Lambda$ to be incorporated in the $T$-matrix. The orange lines show the continuum of negative helicity eigenstates. The blue line in (b) is the positive helicity branch.}
%\label{fig:kspace}
%\end{figure}
\noindent\emph{Note added in proof:}
\newline
After submission of this paper we became aware of several relevant references:

(1) Ref.~\cite{ptok2018} by Ptok, Rodriguez, and Kapcia, referred to in the introduction.

(2) Ref.~\cite{rout2017} by Rout, Maniv, and Dagan ``Link between the Superconducting Dome and Spin-Orbit Interaction in the (111) ${\mathrm{LaAlO}}_{3}/\mathrm{SrTi}{\mathrm{O}}_{3}$ Interface," reporting that the strength of the spin-orbit interaction in that system tracks the magnitude of $T_c$ across the superconducting dome.

(3) Ref.~\cite{stornaiuolo2017} by Stornaiuolo et al. on the same system as Ref.~\cite{rout2017} also suggests such a link.

We point out that our work provides a possible explanation for the observations in Refs.~\cite{rout2017, stornaiuolo2017}, hence suggests that correlated hopping plays an
important role in that system.
\begin{acknowledgments}

We are grateful to the authors of Refs.~\cite{ptok2018, rout2017, stornaiuolo2017} for calling their work to our attention. This work was supported in part by the Natural Sciences and Engineering
Research Council of Canada (NSERC) as well as Alberta Innovates - Technology Futures (AITF).

\end{acknowledgments}

\bibliography{RashbaBCS}

\end{document}